\documentclass[11pt]{article}

\usepackage{amsmath,amsfonts,amsthm,amssymb,latexsym}
\usepackage{bm,bbold}
\usepackage{comment}
\usepackage{graphicx}
\usepackage[margin=1in,letterpaper]{geometry}
\usepackage{indentfirst}
\usepackage{natbib}
\usepackage{soul}
\usepackage{enumitem}
\usepackage{booktabs}
\usepackage{makecell}

\newcommand{\lp}[1]{\left(#1\right)}

\newcommand{\lB}[1]{\left\{#1\right\}}

\newcommand{\lV}[2][]{\left\|#2\right\|_{#1}}

\newcommand{\indicator}[1]{\mathbb{1}_{\lB{#1}}}
\newcommand{\tran}{^\mathsf{T}}

\newcommand{\R}{\mathbb{R}}

\newcommand{\bmp}{\bm{p}}

\newcommand{\bmv}{\bm{v}}
\newcommand{\bmw}{\bm{w}}
\newcommand{\bmx}{\bm{x}}

\newcommand{\bmP}{\bm{P}}

\newcommand{\bmtheta}{\bm{\theta}}
\newcommand{\bmlambda}{\bm{\lambda}}
\newcommand{\bmmu}{\bm{\mu}}

\newcommand{\bmSigma}{\bm{\Sigma}}
\newcommand{\bmOmega}{\bm{\Omega}}
\newcommand{\bmpi}{\bm{\pi}}

\usepackage{caption}
\usepackage{subcaption}
\usepackage{float}     %

\usepackage{array,booktabs}
\usepackage{multirow} 

\newcounter{exhibit}

\renewcommand{\thefigure}{\arabic{exhibit}}
\makeatletter
\let\oldtable\table
\let\endoldtable\endtable
\renewenvironment{table}{%
    \stepcounter{exhibit}%
    \oldtable%
}{\endoldtable}

\let\oldfigure\figure
\let\endoldfigure\endfigure
\renewenvironment{figure}{%
    \stepcounter{exhibit}%
    \oldfigure%
}{\endoldfigure}
\makeatother

\usepackage{hyperref}
\usepackage[table]{xcolor}
\hypersetup{
	colorlinks,
	linkcolor={blue!50!black},
	citecolor={blue!50!black},
	urlcolor={blue!50!black}
}
\usepackage{tablefootnote}

\makeatletter
\newcommand\keyfindingsname{Key Findings}
\makeatother

\title{
Dynamic Factor Allocation Leveraging Regime-Switching Signals
}

\author{
\textbf{Yizhan Shu} \and \textbf{John M. Mulvey}
}

\date{%
\begin{tabular}{rl}
   Current Version:  &   October 17, 2024 \\
   Original Manuscript:  & September 18, 2024
\end{tabular}
}

\begin{document}

\maketitle

\noindent \textbf{Yizhan Shu} is a PhD candidate at Princeton University in the Operations Research and Financial Engineering Department in Princeton, NJ, and is the corresponding author. \\
\href{mailto:yizhans@princeton.edu}{yizhans@princeton.edu}

\vskip 1em

\noindent \textbf{John M. Mulvey} is a professor at Princeton University in the Operations Research and Financial Engineering Department in Princeton, NJ. \\ 
\href{mailto:mulvey@princeton.edu}{mulvey@princeton.edu}

\begin{abstract}

This article explores dynamic factor allocation by analyzing the cyclical performance of factors through regime analysis.
The authors focus on a U.S. equity investment universe comprising seven long-only indices representing the market and six style factors: value, size, momentum, quality, low volatility, and growth. 
Their approach integrates factor-specific regime inferences of each factor index's active performance relative to the market into the Black-Litterman model to construct a fully-invested, long-only multi-factor portfolio.  
First, the authors apply the \emph{sparse jump model} (SJM) to identify bull and bear market regimes for individual factors, using a feature set based on risk and return measures from historical factor active returns, as well as variables reflecting the broader market environment. 
The regimes identified by the SJM exhibit enhanced stability and interpretability compared to traditional methods. 
A hypothetical single-factor long-short strategy is then used to assess these regime inferences and fine-tune hyperparameters, resulting in a positive Sharpe ratio of this strategy across all factors with low correlation among them.
These regime inferences are then incorporated into the Black-Litterman framework to dynamically adjust allocations among the seven indices, with an equally weighted (EW) portfolio serving as the benchmark. 
Empirical results show that the constructed multi-factor portfolio significantly improves the information ratio (IR) relative to the market, raising it from just 0.05 for the EW benchmark to approximately 0.4.
When measured relative to the EW benchmark itself, the dynamic allocation achieves an IR of around 0.4 to 0.5.
The strategy also enhances absolute portfolio performance across key metrics such as the Sharpe ratio and maximum drawdown.
These findings highlight the effectiveness of leveraging regime-switching signals to enhance factor allocation by capitalizing on factor cyclicality.

\end{abstract}


\clearpage

\section{Introduction}

Originating from the Capital Asset Pricing Model (CAPM) \citep{sharpe1964}, which identifies the market as the sole source of risk that earns a premium, factors have gained significant attention as key drivers of the cross-sectional variation in security returns, offering a systematic approach to harvesting additional risk premia. 
Research on factors has expanded beyond common stocks to encompass other assets, such as credit \citep{DeJong2020credit} and real estate \citep{Reid2023}, and even emerging assets like cryptocurrency \citep{liu2022crypto}. 
Factor investing has redefined the traditional approach to asset allocation, shifting the focus from asset classes to factors, in an effort to achieve diversification benefits more effectively \citep{Bass2017}. 
With the development of indexing and exchange-traded funds (ETFs), access to style factors such as value and momentum has extended from actively managed funds to index-tracking smart beta ETFs. 
These rule-based ``passive'' ETFs enable investors to express their active views in a cost-efficient manner \citep{Buetow2024}.

Although many well-known factors have been empirically shown to deliver excess returns over prolonged periods, often spanning decades,\footnote{
For example, the seminal work by \citet{Fama1992} demonstrates the return differences in portfolios formed by sorting based on firm size and book-to-market ratio over an approximately 30-year period from 1963 to 1990.
}
factor performance over shorter horizons is subject to strong cyclicality, with periods of persistent underperformance that can last for years. 
This cyclicality has been recognized by practitioners as early as the 1990s,\footnote{
For example, in a 1992 article \citep{Sharpe1992}, which proposes analyzing equity fund performance based on the value and size factors, William F. Sharpe commented: ``those concerned with these distinctions (in asset characteristics) have focused $\ldots$ on long-run average return differences $\ldots$ Less attention has been paid to likely sources of short-run variability in returns among such groups.''
}  %
and more recently by the significant drawdown experienced in value investing over the past decade \citep{Israel2021}. 
In response, numerous factor timing models have been proposed to predict factor premia \citep{Hodges2017, DiCiurcio2024, Vincenz2024}, utilizing variables such as valuation, sentiment, and macroeconomic indicators. 
However, the challenge of accurately forecasting factor premia has been acknowledged by practitioners \citep{Bender2018}, with some referring to it as the ``siren song'' of these timing efforts \citep{Asness2016}.    %
Nonetheless, there is a consensus that dynamic factor allocation, informed by time-varying market conditions, has the potential to enhance portfolio performance compared to a static allocation strategy.

In this article, we explore the opportunities presented by factor cyclicality through the lens of regime analysis. 
The phenomenon of regime switching has been observed across nearly all asset classes and has gained popularity primarily due to its interpretability: the regimes identified by quantitative models often correspond to real-world events, such as changes in central bank monetary policies \citep{Sims2006}. 
A financial regime is defined by extended and consecutive periods characterized by relatively homogeneous market behavior, while a regime shift signals an abrupt but persistent change in that behavior. 
This framework aligns well with the observed alternation between periods of persistent under- and outperformance in factors. 
Additionally, understanding factor performance through regimes is supported by the business cycle -- a key predictor of factor premia, as demonstrated in \citet{Hodges2017, Polk2020} -- which originally motivated the development of financial regime-switching models \citep{hamilton1989}. 
Previous studies, including \cite{Guidolin2008, Ang2023}, have employed Markov-switching models to perform regime analysis on factor performance, providing valuable insights into the time-varying nature of factor returns.

It is important to clearly distinguish between factor regime analysis and factor timing. 
The key difference lies in the interpretative nature of regime analysis as an unsupervised algorithm, contrasting with the predictive nature of timing strategies. 
While a timing strategy directly attempts to link features known at the end of each period to future factor performance, a regime identification model classifies all historical periods into homogeneous sub-clusters during the training phase, based on characteristics observed at the end of each period. 
No supervision in forecasting future returns is involved. 
Following this in-sample learning process, an online inference algorithm -- coupled with the core assumption of persistence -- is employed in real-time application. 
The goal is that the identified regime at the end of each period will persist in the near future without frequent shifts. 
As \citet{nystrup2015JPM} summarize, the purpose of regime-based strategies is ``not to predict regime shifts or future market movements, but to identify when a regime shift has occurred, and then benefit from the persistence of equilibrium returns and volatilities.''

Our study focuses on the allocation among seven assets: one U.S. market index (serving as a benchmark for the overall equity market) and six long-only U.S. equity smart beta indices representing the style factors of: value, size, momentum, quality, low volatility, and growth.
We aim to maintain a long-only, fully-invested allocation across these seven indices which dynamically adjusts to the identified regimes of factor performance.
This asset universe ensures practicality and ease of implementation, as all indices are accessible through ETFs with relatively low expense ratios. 
Additionally, we incorporate realistic constraints, including transaction costs and a one-day delay in applying the online regime inference.

Our methodology involves conducting regime analysis based on the daily active returns for each of the six factor indices, followed by synthesizing these factor-specific regime inferences using the Black-Litterman framework to incorporate our relative views on each factor's performance. 
For the regime identification model, we highlight the \emph{statistical jump model} (JM), which clusters temporal features while imposing a penalty for each transition in the hidden state sequence. 
This explicit jump penalization allows JMs to moderate the frequency of regime shifts, enhancing stability and robustness against significant noise in market data. Specifically, we implement the \emph{sparse jump model} (SJM) \citep{nystrup2021sparse}, which introduces a layer of feature weighting and selection on top of the JM based on each feature's in-sample clustering effect. 
This extension further improves the applicability of JM in high-dimensional scenarios, where both informative and irrelevant features are present.
Compared to traditional Markov-switching models, recent studies \citep{nystrup2020jump, nystrup2020online, Aydinhan2024, shu2024regime} have demonstrated that JMs achieve higher statistical accuracy and offer better downside risk management.

We input around twenty features into a two-state sparse jump model for each factor index. 
These features include those computed from historical daily factor active returns, as well as others related to the market environment, all of which are relatively easy to retrieve in real time.
The two identified regimes are distinguished by returns, naturally interpreted as bull or bear factor performance. 
For each factor, we apply the regimes inferred in an online fashion to construct a hypothetical long-short strategy that, in essence, involves going long on the factor during anticipated bull markets and shorting the factor otherwise. 
This single-factor long-short strategy mimics the active positions in the Black-Litterman model when a relative view of the active return for a specific factor is inputted. 
Therefore, its performance serves as a financial evaluation of the quality of the regime inference and is used for tuning the hyperparameters in the SJM. 
We achieve a positive Sharpe ratio for the long-short strategy across all factors, with relatively low correlations among them, enabling effective synthesis through the Black-Litterman model.

We adopt the Black-Litterman model in the final stage of portfolio construction due to its practical relevance in portfolio management. 
An equally weighted (EW) portfolio among the seven indices, rebalanced quarterly, serves as the benchmark. 
Since our regime analysis focuses on each factor's active performance relative to the market, the natural choice for view portfolios is six relative portfolios that go long on the factor and short on the market. 
We link the inferred regimes to the expected returns of these view portfolios by calculating the historical average active return across all training periods under the inferred regime.
The posterior expected returns are then input into a standard mean-variance optimization with long-only and fully-invested constraints to determine the optimal allocation.
The confidence level of the views is adjusted to achieve a specified tracking error.

In terms of active\footnote{
Throughout the article, ``active'' refers to returns relative to a benchmark, either the market or the EW portfolio (as described in the previous paragraph). 
In contrast, ``absolute'' refers to returns in excess of the risk-free rate, which is the U.S. Treasury 3-month constant maturity yield.
}
performance, our dynamic allocation strategy significantly improves both the information ratio and max drawdown relative to the market, and achieves an information ratio of approximately 0.4 to 0.5 relative to the EW benchmark.
For absolute performance, our strategy consistently enhances (risk-adjusted) return metrics, such as the Sharpe ratio, as well as risk metrics including volatility and max drawdown, all within a reasonable turnover rate.
These results highlight the benefits of dynamically adjusting factor allocations based on identifying persistent regimes in factor performance.

The outline of the article is as follows. 
We begin with an overview of the factor data used in our analysis in Section \ref{sec:data}. 
The following two sections describe our methodology and empirical results for regime identification and online inference, and the portfolio construction via the Black-Litterman model in Sections \ref{sec:regime} and \ref{sec:portfolio}, respectively.
We conclude with our main findings in the final section.

\section{Asset Universe: Factor Index Data}       \label{sec:data}

The study of optimal allocation among a collection of factors dates back to the early 2000s when \citet{pastor2000, Brennan2001} explored how an investor could exploit market anomalies representing the mispricing of an asset pricing model (\mbox{e.g.}, value and size effects are anomalies with respect to the CAPM). 
Their investment universe typically consisted of a broad market index and several long-short factor portfolios, such as the SMB and HML portfolios constructed by \citet{FAMA1993}. Despite the widespread popularity of these long-short portfolios within academia -- particularly those formed by sorting based on stock characteristics -- they remain far from practical investment opportunities, as they are constructed ``without features or constraints that would make them investable in practice (e.g., limits on position size)'' \citep{Bender2013}, especially for momentum, which inherently involves considerable turnover.

To ensure realism in our study, we adopt an investment universe of seven long-only U.S. equity indices, reflecting the market and six style factors: value, size, momentum, quality, low volatility, and growth. 
The six factor indices represent long-only portfolios with tilts toward specific stock characteristics. 
All seven indices are tracked by actively traded ETFs with assets under management (AUM) in the billions of dollars and are accessible at relatively low expense ratios.\footnote{
For example, the iShares MSCI factor ETFs have an expense ratio of 15 basis points, comparable to many popular cap-weighted ETFs, such as those tracking the Nasdaq-100 or the Nasdaq Composite Index, which have an expense ratio of around 20 basis points.
}
We sourced daily total returns for these indices from Bloomberg, covering the period from 1993 to mid-2024.\footnote{
A detailed description of the indices is provided in Appendix \hyperref[append:data]{A}.
}

The six smart beta indices deviate from market capitalization weighting in their parent index by assigning higher weights to stocks exhibiting targeted characteristics, such as momentum, measured by past stock returns within a specified historical window. 
For many of these indices, an initial screening is conducted so that only a small portion of the stocks -- usually well under half of all stocks in the parent index -- with the most pronounced targeted characteristics receive non-zero allocations. 
This approach results in a stock universe that is quite different from a truly passive cap-weighted index. 
\citet{bender2014alpha} provide a more detailed treatment of these smart beta indices, arguing that they account for more than 80\% of the alphas of all active managers.

\begin{table}[p]
    \centering
    \begin{tabular}{l *{6}{r} }
    \toprule
    & Value & Size & Momentum & Quality & Low Vol & Growth \\
    \midrule
    Active Return (\%) & 0.35 & 0.11 & 2.98 & 0.99 & $-$1.03 & 0.78 \\
    Information Ratio & 0.06 & 0.02 & 0.39 & 0.29 & -- & 0.14 \\
    Max Drawdown (\%) & $-$44.8 & $-$40.4 & $-$30.9 & $-$12.1 & $-$40.6 & $-$42.5 \\
    \midrule
    $\alpha$ (\%) & 0.36 & 0.13 & 3.10 & 1.31 & 0.98 & 0.29 \\
    $t$-stat & 0.35 & 0.15 & 2.27 & 2.17 & 1.21 & 0.31 \\
    $\beta$ & 1.00 & 1.00 & 0.99 & 0.97 & 0.78 & 1.05 \\
    \bottomrule
    \end{tabular}

    \vspace{2mm}
    
    \parbox{\textwidth}{\footnotesize 
    Notes: 
    $\alpha$ and $\beta$ are estimated from univariate ordinary least squares (OLS) regressions against the market, using all available data.
    ``(\%)'' indicates that the displayed numbers are in percentage points. 
    All values are annualized where applicable.
    For instance, the annualized active return for the value factor is 0.35\%.
    }

    \caption{Active Performance Analysis Relative to the Market Index for Six Factor Indices, January 4, 1993 -- June 28, 2024}
    \label{tab:index act perf}
\end{table}

\begin{figure}[p]
    \centering
    \includegraphics[width=\textwidth]{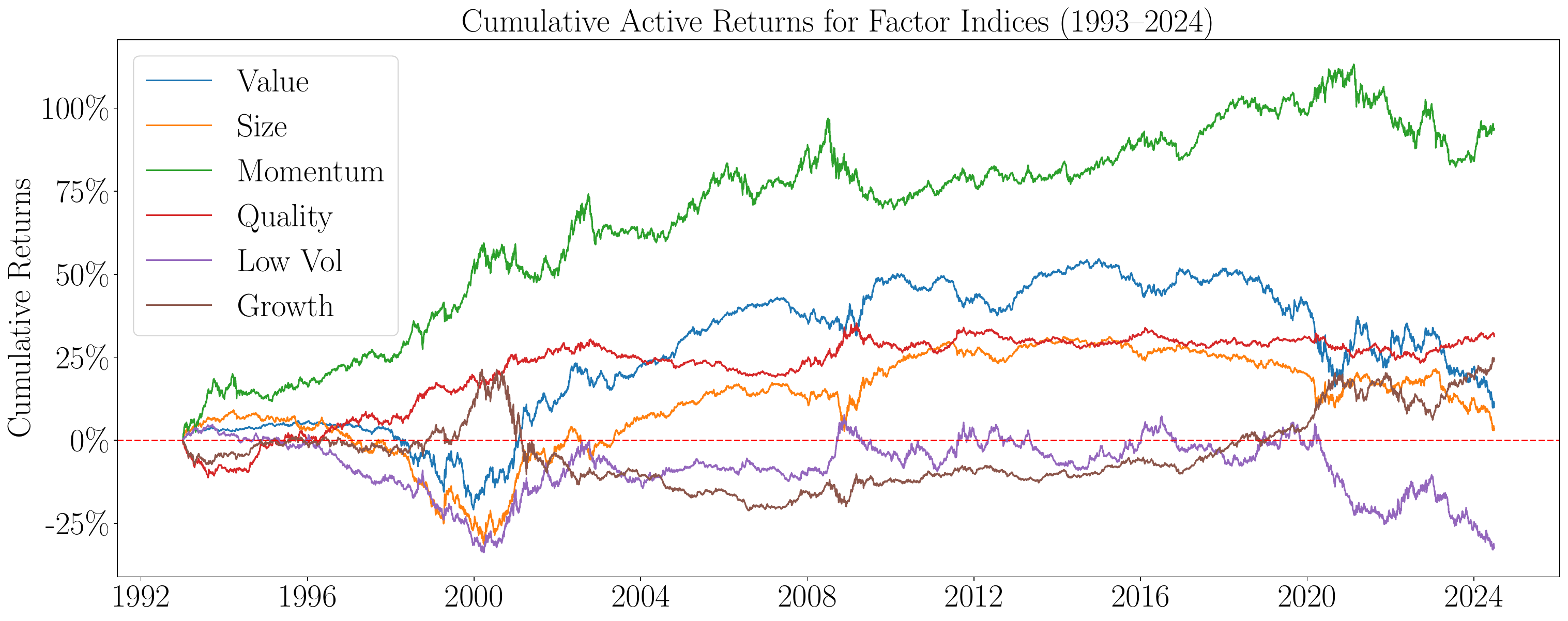}
    
    \includegraphics[width=\textwidth]{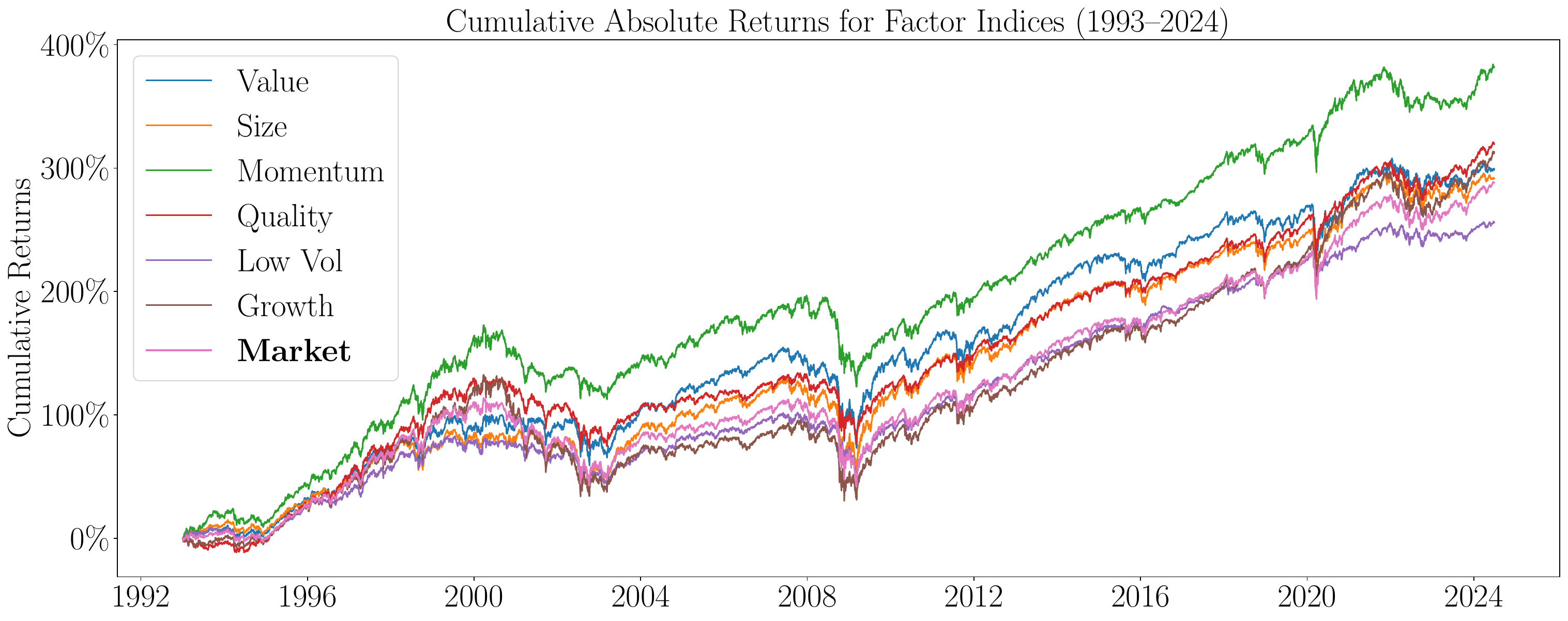}
    
        \vspace{2mm}

    \parbox{\textwidth}{\footnotesize Notes: Active returns are measured relative to the market index return, while absolute returns are measured relative to the risk-free rate.}

    \caption{Cumulative Active and Absolute Returns for All Factor Indices, January 4, 1993 -- June 28, 2024}
    \label{fig:cumret}
\end{figure}

Despite the long-only index construction, we assess factor performance on a relative basis by computing factor active returns, defined as the factor index return minus the market index return.\footnote{
Our use of active returns follows the approach in \citet{Hodges2017}. 
Alternatively, employing residual returns that account for betas could potentially address the inherent low-beta effect in the low volatility factor index, as discussed later, and can also be easily integrated into the Black-Litterman model.
} 
This relative perspective aligns with the original discovery of factors as the outperformance of one group of stocks versus another, and also helps reduce correlation when performing regime analysis on individual factors. 
Although factor indices involve weight tilts and stock screening, they still maintain a considerable long-run beta to the market, typically ranging from 0.75 to 1.1 (see Exhibit~\ref{tab:index act perf}), resulting in a high correlation among absolute factor returns.\footnote{
Our data reveals that the highest correlation of factor active returns is 54\% between value and size, and $-$52\% between value and growth; however, all correlations between factor returns in excess of the risk-free rate exceed 85\%. 
These figures are consistent with those reported in \citet{Hodges2017, Polk2020, DiCiurcio2024}. 
Although the last article in this list does not explicitly work with active returns, it attempts to forecast the winning factor, which still operates on a comparative rather than absolute basis.
}
This reduction in correlation is evident from the cumulative active and absolute return curves for all factor indices shown in Exhibit~\ref{fig:cumret}. 
By removing the common market component, we reveal distinct patterns for various factors, contrasting with the highly correlated growth trends seen in the absolute return curves.
It is worth noting that although we assess factor performance on a relative basis, the final multi-factor portfolio remains a long-only allocation across the seven indices, with no shorting allowed, to ensure practicality.

In the active performance table for all factors presented in Exhibit~\ref{tab:index act perf}, we observe positive active returns for all factors except low volatility, which is attributed to its inherently low beta. 
After adjusting for market risk, all factors display a positive alpha intercept, reflecting their design to achieve positive risk premia, although most of these are not statistically significant.

Exhibit~\ref{fig:cumret} further highlights the increased difficulty of performing regime analysis based on active factor performance compared to absolute performance, as analyzed in \citet{Bosancic2024}. 
This challenge arises from the lack of a strong trend in active returns, in contrast to the clear upward and downward trends observed in absolute return curves. 
Indeed, factor active returns are only a few percentage points per annum, compared to the approximately 8--10\% expected from a market index. 
Moreover, the sharp drawdowns experienced by a fully invested index during major market crashes -- such as the dot-com bubble and the global financial crisis (GFC), whose accurate detection has been shown by \citet{mulvey2016, shu2024regime} to be effective in mitigating downside risk -- are not prevalent in the active return plots.\footnote{
This is evidenced by both the magnitude and duration of the max drawdown (MDD) seen in Exhibits \mbox{\ref{tab:index act perf}   and \ref{fig:cumret}}. 
While a market index often experiences an MDD of more than $-$50\% within one to two years, factor active returns typically undergo an MDD of approximately $-$40\% over a much longer period. 
For example, value's MDD relative to the market began at the start of 2015 and has persisted until recently, lasting nearly a decade. 
Even momentum, known for ``momentum crashes'' \citep{DANIEL2016}, experienced its relative MDD between February 2021 and July 2023, spanning over two years. 
Its post-GFC crash lasted around a year and a half, from mid-2008 to the end of 2009, comparable to the market, albeit with a much milder magnitude of $-$27.5\%.
}
While most previous regime-based strategies, such as those in \citet{bulla2011strategy, nystrup2015JPM, Nystrup2018JPM}, focus on identifying the regimes of a long-only equity index to enhance asset allocation performance, in this article, we extend this approach by demonstrating the use of regime-switching signals based on active performance.

Regarding factor correlations, as discussed in \citet{Bender2016}, a ``bottom-up'' approach, which constructs a multi-factor portfolio at the security level by incorporating all factor characteristics simultaneously, offers greater potential to exploit interaction effects among factors compared to our ``combination approach'', which directly allocates among a selected number of pre-constructed factor portfolios. 
The bottom-up approach is sometimes preferred by asset managers and ETF providers. 
However, our allocation approach is inherently easier to implement and is more accessible to investors who lack the resources to perform nuanced stock-level analysis.

\section{Regime Identification and Online Inference}         \label{sec:regime}

\subsection{Statistical Jump Models}

Our study employs the sparse version of the recently developed statistical jump model (JM) as the regime identification model.
While machine learning clustering algorithms -- such as the $k$-means clustering used to detect market risk regimes \citep{DiCiurcio2024} -- have gained popularity in financial applications, JMs are particularly well-suited to time series data due to their incorporation of temporal information through an explicit jump penalty for each transition in the hidden state sequence. 
This approach acknowledges the inherent persistence in financial regimes and enhances the interpretability of the identified regimes, as market participants typically do not expect frequent regime shifts.
\citet{Bosancic2024} recently demonstrated the successful application of JMs in conducting regime analysis for long-only factor portfolios.

Given a series of high-dimensional standardized features $\bmx_0,\ldots,\bmx_{T-1}$ (detailed in the next subsection), we estimate a two-state JM ($K=2$) by solving the following optimization problem:
\begin{equation}\label{eq:jumpObj}
\min_{\bmtheta_0,\ldots,\bmtheta_{K-1}, s_0,\ldots, s_{T-1}}\quad\sum_{t = 0}^{T-1} \frac12\lV[2]{\bmx_t-\bmtheta_{s_t}}^2 +\lambda\sum_{t = 1}^{T-1} \indicator{s_{t-1} \neq s_{t}} \,.
\end{equation}
Here, the optimization variables include the unobserved state sequence $s_0,\ldots, s_{T-1}$, which assigns each period to one of the $K$ states, and $K$ centroids $\bmtheta_0,\ldots,\bmtheta_{K-1}$, representing each state. 
The dissimilarity between the feature vector $\bmx_t$ and its assigned centroid $\bmtheta_{s_t}$ is quantified by the squared $\ell_2$-distance. 
From the optimal state sequence, we derive the transition probability matrix and calculate metrics such as return and volatility for each state.\footnote{
An implementation of a collection of jump models is available on the first author’s GitHub page (\url{https://github.com/Yizhan-Oliver-Shu/jump-models}).
For a detailed explanation of the coordinate descent algorithm used to solve the optimization problem, see \citet{nystrup2020jump}.
}
The JM’s objective function balances the partitioning of all training periods into homogeneous sub-clusters with prior beliefs about the frequency of regime shifts. 
The jump penalty $\lambda \ge 0$ moderates the fixed-cost regularization term incurred whenever there is a jump between consecutive state variables (\mbox{i.e.}, when $s_{t-1} \neq s_{t}$). 
This penalty controls the persistence level of the state sequence and is tuned as a hyperparameter, often via cross-validation. 
With a zero jump penalty, the model reduces to a $k$-means clustering algorithm, ignoring temporal information; as $\lambda$ increases, state transitions become less frequent, potentially grouping all data points into a single cluster if $\lambda$ is set high enough.

To further enhance the JM, we implement the sparse jump model (SJM) \citep{nystrup2021sparse}, which adds a layer of feature weighting based on each feature’s clustering effect, measured by the reduction in variance along its dimension. 
Features that lead to greater variance reduction are assigned higher weights, while an $\ell_1$-norm constraint on the weight vector ensures that uninformative features with poor clustering effects receive a weight of zero and are excluded from consideration. 
The upper bound $\kappa$ on the $\ell_1$-norm serves as an additional hyperparameter, controlling the number of features included. 
The SJM estimation alternates between fitting the JM using weighted features and recalculating the feature weights based on the current JM fit until convergence is reached. 
This extension addresses the original JM’s limitation of assigning equal weights to all features, improving its ability to extract useful information from high-dimensional feature sets, which are increasingly common in modern financial data.

The suite of JMs offers several advantages over traditional Markov-switching models, such as hidden Markov models (HMMs), in financial applications. 
Their primary strength lies in enhanced robustness and stability, achieved through explicit jump penalization and a likelihood-free approach. 
Recent studies, including \citet{nystrup2020jump, nystrup2020online, Aydinhan2024}, highlight the susceptibility of HMMs to mis-estimation due to issues like imbalanced clusters (\mbox{e.g.}, the small proportion of bear markets) and insufficient data, as well as the mis-specification of likelihood functions. 
These studies, through stylized simulations and real-data tests, demonstrate the improved statistical accuracy of JMs and their continuous version over HMMs. \citet{shu2024regime} further show that this improved accuracy can translate into better investment strategy performance.
Additionally, JMs naturally accommodate high-dimensional feature sets, unlike HMMs, which are typically applied to single-dimensional return series. 
\citet{nystrup2021sparse} highlight the advantages of SJMs over high-dimensional Markov-switching models, while \citet{Bosancic2024} demonstrate the flexibility of JMs in accommodating regime-dependent features.

\subsubsection{Features}

Exhibit~\ref{tab:features} provides a detailed description of all features inputted into the SJM, categorized into two types: factor-specific features derived from historical factor active returns and general market-environment features. 
For the first type, when performing regime analysis on each individual factor, we calculate a collection of risk and return measures based on the past active returns of that factor. 
The return measures include the exponentially weighted moving average (EWMA) active return, along with three technical indicators -- relative strength index (RSI), stochastic oscillator \%$K$, and Moving Average Convergence/Divergence (MACD) -- all computed over three window lengths corresponding to 1.5 weeks, one month, and three months. 
These return features aim to capture the time-series momentum effect substantiated in \citet{Gupta2019} for factor returns, and have been applied in various timing models such as those in \citet{Zakamulin2014, Hodges2017}. 
Their application in identifying regimes in cryptocurrency markets via SJMs is demonstrated in \citet{cortese2023crypto}. 
For risk measures, we compute downside deviation, focusing on negative active returns, and active market beta, both using an exponentially weighted calculation with a medium window length.
It is important to note that these features are based on past returns and thus lag the regime shift identified ex-post. 
It is worth reiterating that regime analysis does not aim to predict regime shifts but rather to benefit from accurately identifying when they have occurred, allowing for allocation adjustments and gaining an advantage over sticking to a static strategy.

\begin{table}[tb]
\centering
\begin{tabular}{llcc}
\toprule
\makecell[c]{\textbf{Feature} \\ \textbf{Name}} & \textbf{Transformation}  & \makecell[c]{\textbf{Window} \\ \textbf{Lengths}}   & \textbf{Abbreviation} \\
\midrule
\multicolumn{4}{l}{\textbf{Category: Factor-Specific Features}} \\
\midrule
Active Return                  & EWMA & 8, 21, 63         & $r^{\text{factor}}_w$\\
Relative Strength Index         & -- & 8, 21, 63          & $\text{RSI}_w$\\
Stochastic Oscillator \%$K$    & --  & 8, 21, 63          & $\%K_{w}$ \\
MACD                            & -- &  (8, 21), (21, 63)  & $\text{MACD}_{s, l}$\\
Downside Deviation              & log  & 21                 & $\log\lp{\text{DD}_w}$\\
Active Market Beta                     & --   & 21                  & $\beta_w$\\
\midrule
\multicolumn{4}{l}{\textbf{Category: Market-Environment Features}} \\
\midrule
 Market Return                 & EWMA            &  21  &  $r^{\text{mkt}}_w$\\
VIX                            & log, diff, EWMA & 21   &  $r^{\text{VIX}}_w$ \\
2Y Yield                       & diff, EWMA      &  21   &   $\text{2Y}\_\text{diff}_w$\\
10Y Minus 2Y Yield             & diff, EWMA      &  21   &    $\text{10Y2Y}\_\text{diff}_w$\\
\bottomrule
\end{tabular}

    \vspace{2mm}

    \parbox{\textwidth}{\footnotesize Notes: All window lengths are measured in trading days and refer to the effective window size in the exponentially weighted moving (EWM) computation, except for \%$K$, where they refer to the rolling window length. 
    ``EWMA'' stands for EWM Average. 
    ``diff'' stands for the difference between consecutive values.
    ``MACD'' stands for Moving Average Convergence/Divergence.
    All factor-specific features are computed based on factor active returns.
    Transformations are applied in sequence; for example, the VIX feature is calculated as the EWMA of the log-returns of the VIX index, where positive/negative values indicate an increasing/decreasing volatility environment. 
    Raw data for the last three market-environment features are sourced from the FRED Database.    
     }

\caption{Overview of Features Inputted into the Sparse Jump Model}     
\label{tab:features}
\end{table}

The second type of features focuses on the general market environment, measuring market returns, the VIX index, the 2-year Treasury constant maturity yield as an interest rate gauge, and the yield curve slope measured by the difference between the 10-year and 2-year yields.
Appropriate transformations, such as log-differences and exponential smoothing, are applied to these features. 
The market environment naturally plays a critical role in driving factor performance. 
Specifically, \citet{Hodges2017} found that the business cycle is a highly effective predictor of factor premia, with a stylized pattern where economic slowdowns favor defensive strategies like quality and low volatility, while value, with its higher financial leverage, tends to plunge -- this pattern reverses during periods of economic expansion. 
We use the aforementioned variables, easily sourced from the FRED Database, as proxies for the general market environment. 
These variables are based on trading markets, updated daily, and are readily accessible in real time without delay.

\subsubsection{An Example of SJM Fitting}

\begin{figure}[tbp]
    \centering
    
    \includegraphics[width=\textwidth]{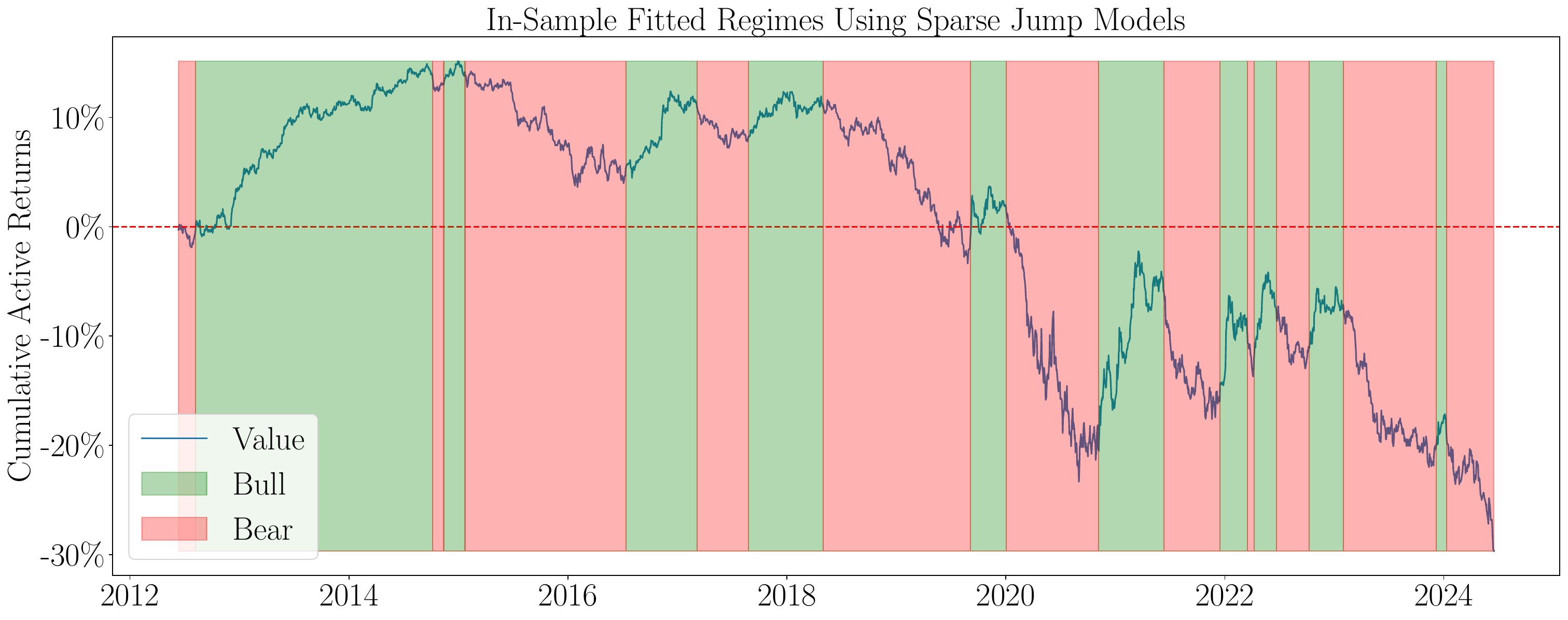}

    \includegraphics[width=\textwidth]{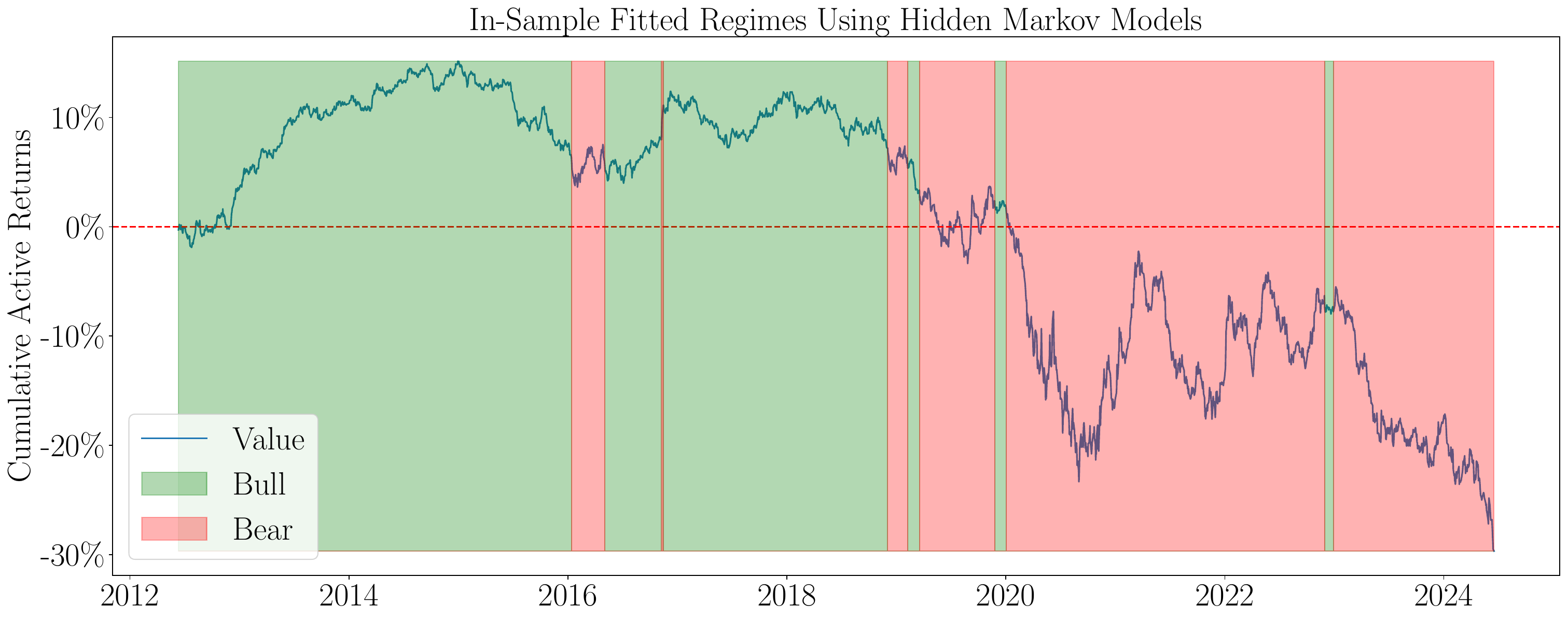}

    \includegraphics[width=\textwidth]{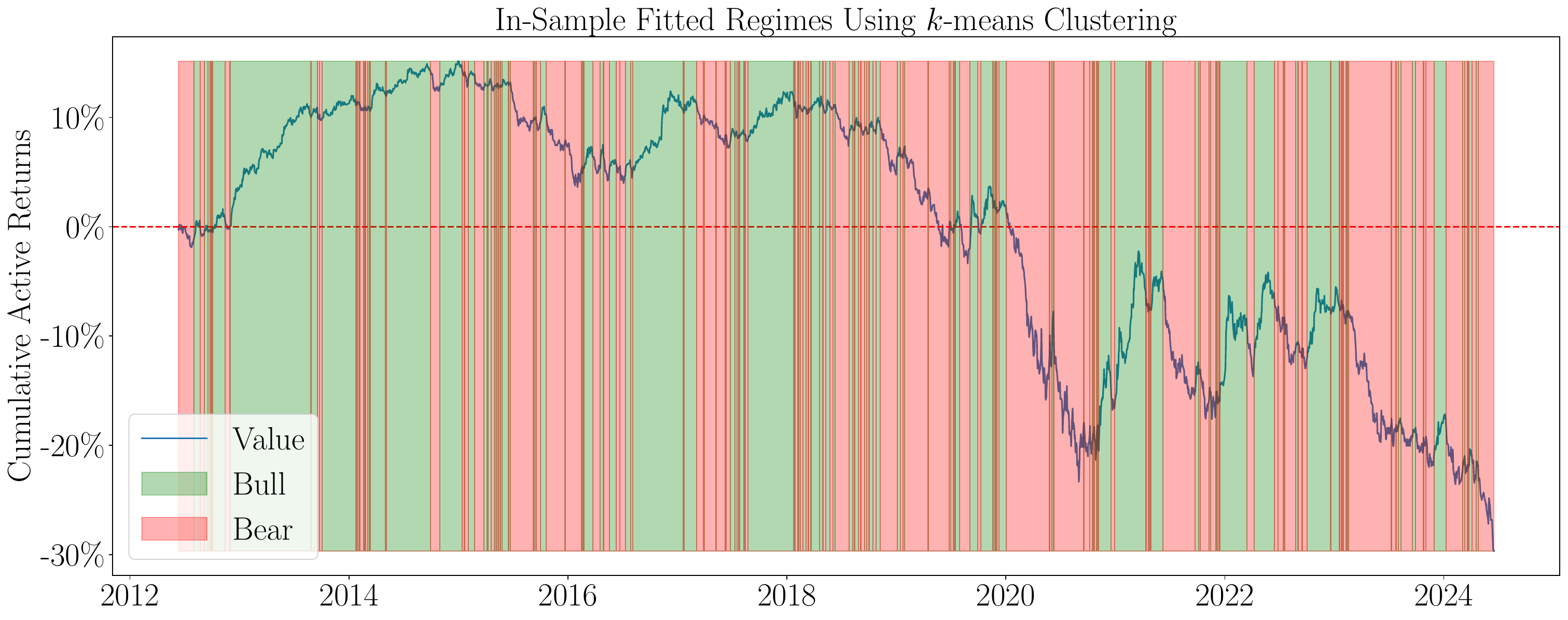}

    \vspace{2mm}

    \parbox{\textwidth}{\footnotesize 
    Notes: The blue curve represents the cumulative active returns of the value factor index; the green and red shaded areas represent the in-sample fitted bull and bear regimes, respectively, distinguished by the cumulative active returns within each state.
    Sparse jump models and $k$-means clustering use the features listed in Exhibit \ref{tab:features}; hidden Markov models use factor active returns as input. 
    The SJM is configured with typical hyperparameters: $\lambda=50.0$ and $\kappa^2=9.5$.
     }

    \caption{In-Sample Fitted Regimes Using Sparse Jump Models, Hidden Markov Models, and $k$-means Clustering (from top to bottom) for the Value Factor, June 2012 -- June 2024}

    \label{fig:value example}
\end{figure}

To better illustrate the mechanism of SJMs and facilitate comparison with baseline models, we present an example fitting on the value factor over a recent 12-year training window, spanning from June 2012 to June 2024. 
Exhibit~\ref{fig:value example} displays the in-sample fitted regimes from three models: SJMs, HMMs, and $k$-means clustering, while Exhibit~\ref{tab:value example weight} shows the feature weights and corresponding optimal centroid values under the two market states for each feature dimension in the SJM fitting. 
We distinguish between bull and bear market states based on the cumulative active return for each state across all training periods.
The SJM and $k$-means clustering are estimated using the features described in Exhibit~\ref{tab:features}, with the SJM hyperparameters set to typical values,\footnote{
In our empirical study, the SJM hyperparameters are tuned using a time series cross-validation approach to ensure practicality, as detailed in the following subsection.
} while the HMM is fitted using daily active returns as input.

From the figures, we observe that the SJM accurately captures the persistent periods of out- and underperformance of the value factor, including the rebound after the GFC, the recent plunge following the COVID-19 market crash, and the high-interest-rate environment from 2023.
A few rebounds between these bear markets are also identified. 
This level of persistence contrasts sharply with $k$-means clustering, which does not account for temporal consistency and results in an unrealistically large number of regime shifts.
The SJM can be seen as an effective smoothing of the $k$-means fitting, enhanced by the feature weighting procedure. 
As for HMMs, the balanced nature of bull and bear markets in factor performance makes the issue of lacking persistence less pronounced than in applications to market indices, as illustrated, for example, in Figure 6 of \citet{shu2024regime}. 
Nonetheless, a few short-lived regime shifts are still present, and if these are overlooked, the model primarily indicates a bull market before 2019 and a bear market afterward, neglecting much of the market dynamics and potential opportunities.

In terms of feature weights, Exhibit~\ref{tab:value example weight} shows that the factor-specific return features with the longest window sizes receive the highest weights. 
The quarterly horizon aligns with the timeframe market participants typically use to assess factor performance. 
The four market-environment features receive very little weight, which may underestimate the influence of the market environment on factor performance. 
Exploring more informative market features could potentially enhance the performance of the SJM.

\begin{table}[tb]
    \centering
\begin{tabular}{lrrr}
\toprule
 & Weight (\%) & Bull & Bear \\
\midrule
$\text{RSI}_{63}$               & 11.24 & 55.50 & 44.19 \\
$\%K_{63}$                      & 10.92 & 76.44 & 22.82 \\
$r^{\text{factor}}_{63}$ (\%)   & 10.47 & 6.70 & $-$9.35 \\
$\text{MACD}_{8, 21}$           & 9.81 & 0.53 & $-$0.64 \\
$\text{RSI}_{21}$               & 9.78 & 57.82 & 42.10 \\
$\text{MACD}_{21, 63}$          & 9.35 & 0.84 & $-$1.32 \\
$r^{\text{factor}}_{21}$ (\%)   & 9.08 & 10.66 & $-$12.76 \\
$\text{RSI}_{8}$                & 6.61 & 58.08 & 41.47 \\
$\%K_{21}$                      & 6.19 & 66.14 & 33.22 \\
$r^{\text{factor}}_{8}$ (\%)    & 5.81 & 11.83 & $-$13.89 \\
$\text{DD}_{21}$ (\%)           & 5.44 & 4.66 & 6.98 \\
$\%K_{8}$                       & 3.18 & 59.19 & 38.59 \\
$\text{2Y}\_\text{diff}_{21}$ (\%)      & 1.17 & 0.69 & 0.12 \\
$\text{10Y2Y}\_\text{diff}_{21}$ (\%)   & 0.96 & 0.04 & $-$0.33 \\
$\beta_{21}$                    & 0.00 & -- & -- \\
$r^{\text{mkt}}_{21}$           & 0.00 & -- & -- \\
$r^{\text{VIX}}_{21}$           & 0.00 & -- & -- \\
\bottomrule
\end{tabular}

    \vspace{2mm}

    \parbox{\textwidth}{\footnotesize 
    Notes: See Exhibit~\ref{tab:features} for a detailed description of all features. 
    The SJM is configured with typical hyperparameters $\lambda=50.0$ and $\kappa^2=9.5$.
    For the downside deviation (DD) feature, the average of the original DD values (multiplied by $\sqrt{2}$ to align with the volatility computation) within each state is displayed, although the logarithm of the DD value is used as an input feature in the SJM. 
    All values are annualized where applicable. 
     }

    \caption{Feature Weights and Optimal Centroid Values Under Bull and Bear Market Regimes for the Sparse Jump Model Fitting for the Value Factor, June 2012 -- June 2024}

    \label{tab:value example weight}
\end{table}

Regarding optimal centroids, these values align well with expectations: in bull markets identified for the value factor, return features exceed baseline values, with positive EWMA active returns, RSI and \%$K$ values above 50, and positive MACD -- all indicating strong price momentum -- and a lower risk measure of downside deviation. 
The opposite is true for bear markets. 
The differences between the centroid values under the two states are also quite discernible.

\subsubsection{Online Inference}

All previous exposition focuses on the in-sample training of SJMs, where the estimated hidden state sequence $s_0, \ldots, s_{T-1}$ is derived using all market data up to the end of period $T-1$.
However, for real-time applications, we must infer the state $s_t$ based solely on data available at the end of the current period, without relying on future information.
This distinction is commonly referred to as \emph{smoothing} and \emph{filtering} in signal processing: smoothing uses both past and future data points to estimate values, while filtering relies only on current and past data, making it suitable for real-time applications.

In the case of JMs, after estimating the optimal centroids and state sequence during the training phase, we must make real-time inferences when a new feature vector $\bmx_T$ becomes available at the end of day $T$. 
For investment purposes, following the approach of \citet{nystrup2015JPM, Nystrup2018JPM}, we assume that this inferred regime $s_T$ can only be applied on day $T+2$ with a one-day delay -- meaning that we estimate $s_T$ at the end of day $T$ and rebalance the portfolio accordingly at the end of the next day, $T+1$. 
This method of applying online inferred regimes to adjust allocations relies heavily on the assumption of regime persistence, whereby the regime identified for period $T$ is expected to persist in the near future.
Consequently, this approach places a high demand on the stability of the inferred regime sequence.
Admittedly, a rapidly flipping regime-switching signal (\mbox{e.g.}, the regimes identified in Exhibit \ref{fig:value example} for $k$-means clustering) is unlikely to remain effective given the time discrepancy between $T$ and $T+2$.

A basic $k$-means-style online inference involves directly assigning the feature vector $\bmx_T$ to the nearest centroid, ignoring temporal information.
To enhance regime persistence, we employ the online inference algorithm outlined in \citet{nystrup2020online}, which efficiently incorporates historical data. 
The core idea is to solve the optimization problem \eqref{eq:jumpObj} in JMs over the state sequence within a lookback window, while keeping the centroids fixed at their previously estimated values, and then extract the last optimal state as the online inference. 
This online algorithm performs the task for a sequence of consecutive inferences in linear time.

Despite this improvement in persistence, it is important to note that the number of regime shifts in real-time inference remains higher -- usually twice as frequent -- than in the in-sample training.
This discrepancy arises because both the first and last few states in a window suffer from the highest estimation error, as shown in Figure 1 of \citet{bulla2011strategy}, due to the lack of past or future data context. 
By analogy, pinpointing the exact onset of a bear market shift (as identified in hindsight) in real-time is challenging; both individuals and quantitative models typically require several consecutive weeks of negative returns to confirm the shift.

An alternative approach to utilizing JMs for generating regime forecasts is to train a separate algorithm that predicts the in-sample identified regime labels $s_t$ using another feature vector~$\tilde\bmx_{t-1}$, which is strictly available beforehand, as illustrated in \citet{shu2024AA}.

\subsection{Single-Factor Long-Short Strategy}

The previous discussion on training the SJM and executing online inference assumes a fixed set of hyperparameters. 
To optimize these parameters, we evaluate the quality of the online inferred regimes by constructing a hypothetical long-short strategy, where we generally go long on the factor during anticipated bull markets and short during bear markets.
\citet{Cortese2024GIC} proposes an alternative method for model selection in JMs using generalized information criteria (GIC).

The construction of this factor-specific long-short strategy closely relates to the active weights derived from a relative view on the active return for a specific factor in the Black-Litterman model, while neglecting the influence of other views (details in Appendix~\hyperref[append:BL]{B}).
To implement this strategy, we first associate the inferred regimes for a specific factor with the historical average active return across all training periods under the same regime. 
This historical average serves as the expected active return for the factor, which also functions as the expected return for the relative view portfolio in the Black-Litterman model. 
It is not intended to be an accurate active return forecast but rather a convenient way to represent the inferred regimes without explicit human interpretation.\footnote{
There are situations where both regimes display positive average returns, so the expectation of a clear bull/bear regime distinction may not hold.
}

When the expected return exceeds 5\% per annum, the strategy invests 100\% in a relative portfolio by going long 100\% on the factor and shorting 100\% of the market. Conversely, when the expected return falls below $-$5\%, the strategy shorts the factor 100\% and longs the market 100\%. 
If the expected returns fall within the range of $-$5\% to 5\%, we invest in this relative portfolio using a linear weighting.
Thus, this strategy invests in a relative portfolio reflecting the factor's active returns, with the position size proportional to the expected factor active return, capped at $\pm$5\%. 
For performance evaluation, a transaction cost of 5 basis points for both buying and selling is consistently applied. 
A radical regime shift from fully bullish to bearish would involve a one-way 200\% turnover (100\% to $-$100\% for the factor and the opposite for the market position).

The Sharpe ratio of this factor-specific long-short strategy serves as the financial evaluation of the online inferred regimes, and we adjust the SJM hyperparameters accordingly using a time-series cross-validation approach. 
To avoid optimizing over a large grid, we tune the hyperparameters for each factor individually. 
Our expanding training window spans a minimum of 8 years and a maximum of 12 years.
For each set of hyperparameters, we refit the SJM monthly and perform online inference during the intervals between refittings. 
The validation window spans 6 years and rolls forward every six months, during which we select the hyperparameter combination that yields the highest Sharpe ratio for the long-short strategy, applying these values for the next six months. 
Our testing period begins in 2007.
It is worth noting that this single-factor long-short strategy is intended solely for evaluating the inferred regimes and not for practical trading, which is why we call it ``hypothetical.''
The final multi-factor portfolio involves a long-only allocation among the seven indices.

\begin{table}[tb]
    \centering
\begin{tabular}{lrrrrrr}
\toprule
\multicolumn{7}{l}{\textbf{Panel A: Single-Factor Long-Short Strategy Performance}} \\ 
\midrule
 & Value & Size & Momentum & Quality & Low Vol & Growth \\
\midrule
Sharpe Ratio & 0.39 & 0.20 & 0.16 & 0.21 & 0.30 & 0.37 \\
\#(Shifts) & 3.16 & 2.57 & 3.66 & 0.64 & 1.78 & 1.31 \\
\bottomrule

\\

\toprule
\multicolumn{7}{l}{\textbf{Panel B: Correlation of the Single-Factor Long-Short Strategy Returns}} \\ 
\midrule
 & Value & Size & Momentum & Quality & Low Vol & Growth \\
\midrule
Value &  & 0.37 & 0.20 & 0.07 & 0.17 & 0.48 \\
Size &  &  & 0.11 & 0.05 & 0.12 & 0.29 \\
Momentum &  &  &  & $-$0.01 & 0.10 & 0.23 \\
Quality &  &  &  &  & 0.21 & 0.11 \\
Low Vol &  &  &  &  &  & 0.28 \\
Growth &  &  &  &  &  &  \\
\bottomrule
\end{tabular}

    \vspace{2mm}

    \parbox{\textwidth}{\footnotesize 
    Notes: 
    The return of this single-factor long-short strategy generally equals the factor's active return during bull regimes and its negative active return during bear regimes, before accounting for transaction costs.  
    This hypothetical strategy is intended solely for evaluating the online inferred regimes and not for practical trading.  
    ``\#(Shifts)'' stands for the number of regime shifts per year.
    A transaction cost of 5 basis points is applied for both buying and selling.  
    All values are annualized where applicable. 
     }

    \caption{Performance of the Hypothetical Single-Factor Long-Short Strategy and Correlation of Strategy Returns Among Factors Over the Test Period of 2007--2024}
    
    \label{tab:single factor perf}
\end{table}

\begin{figure}[tbp]
    \centering
    \includegraphics[width=\textwidth]{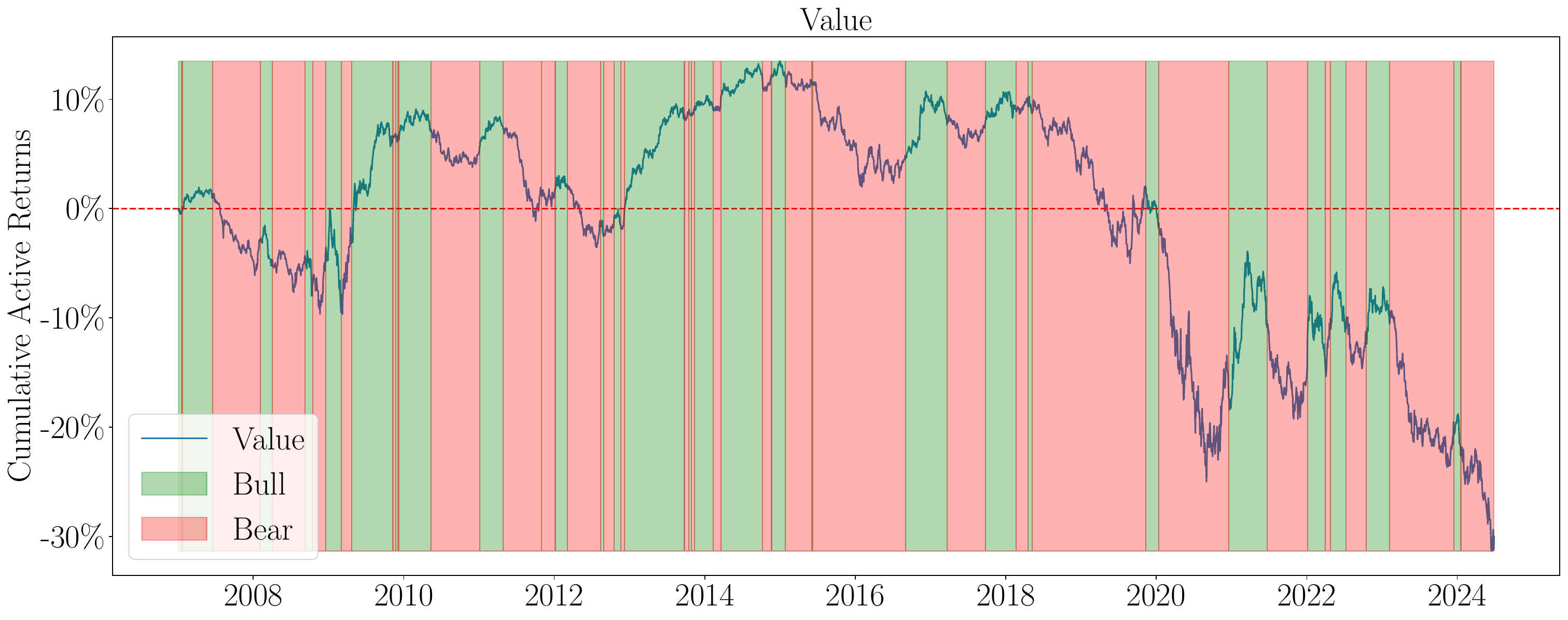}

    \includegraphics[width=\textwidth]{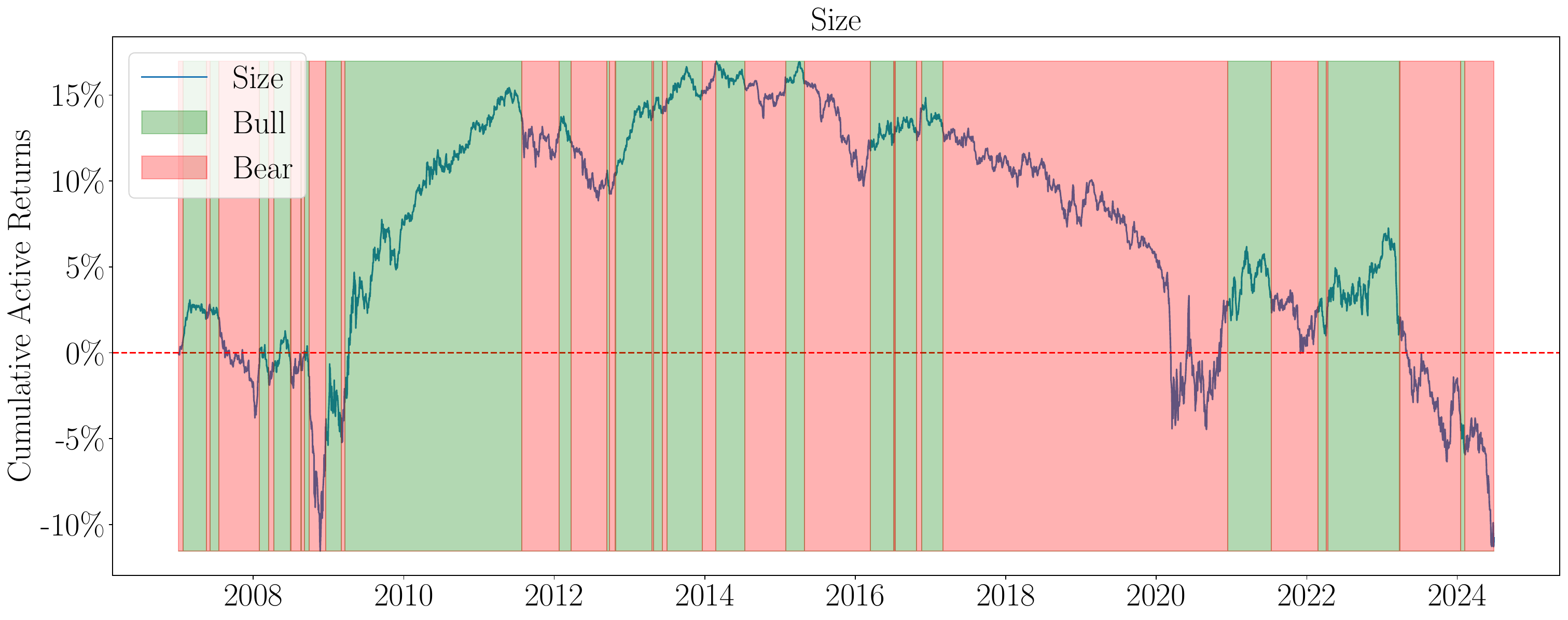}

    \includegraphics[width=\textwidth]{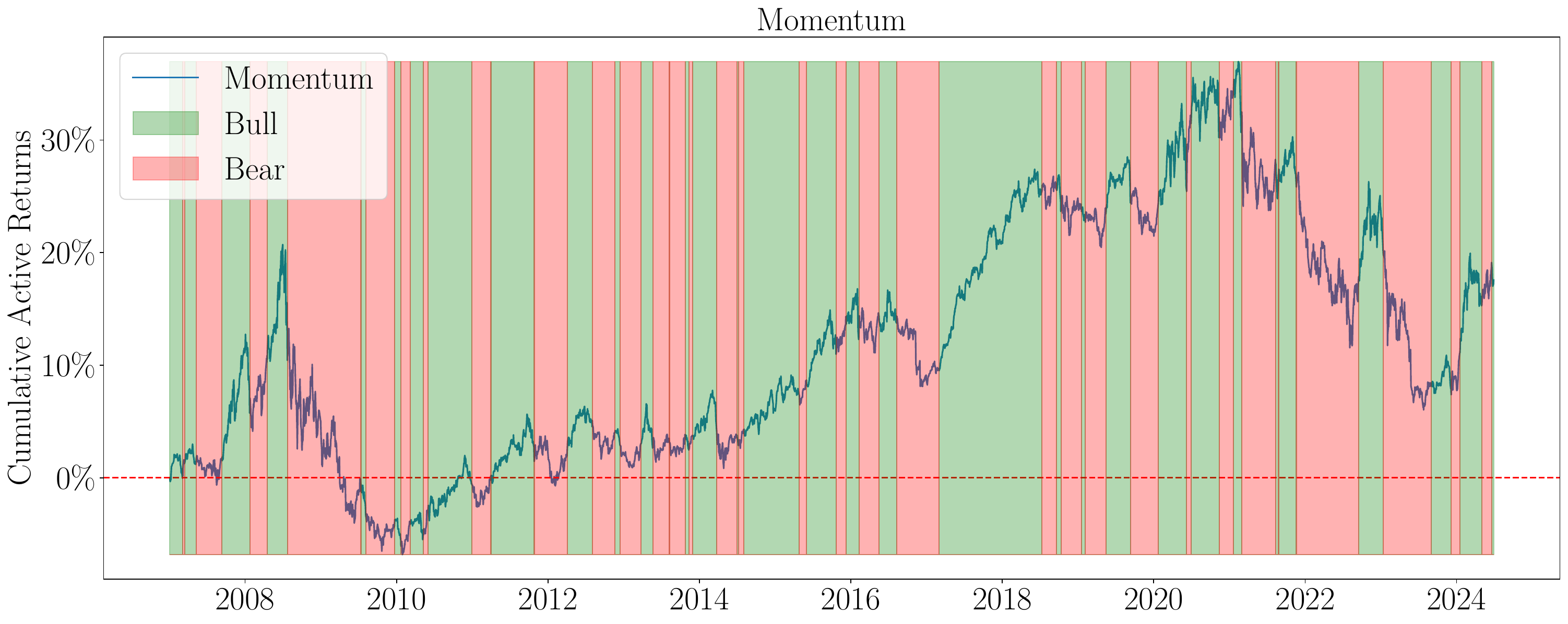}

    \vspace{2mm}

    \parbox{\textwidth}{\footnotesize
    Notes: The blue curve represents the cumulative factor active returns; the green and red shaded areas represent the online inferred bull and bear regimes after hyperparameter tuning, respectively, defined by whether the expected factor active return is positive or negative.
     }

    \caption{Online Inferred Regimes by the Sparse Jump Model (After Hyperparameter Tuning) Over the Test Period of 2007--2024 for Three Factors: Value, Size, and Momentum}
    \label{fig:oos regimes 1}
\end{figure}

\begin{figure}[tbp]
    \centering
    
    \addtocounter{exhibit}{-1}  %
    \renewcommand{\thefigure}{\arabic{exhibit} (Continued)} %
    
    \includegraphics[width=\textwidth]{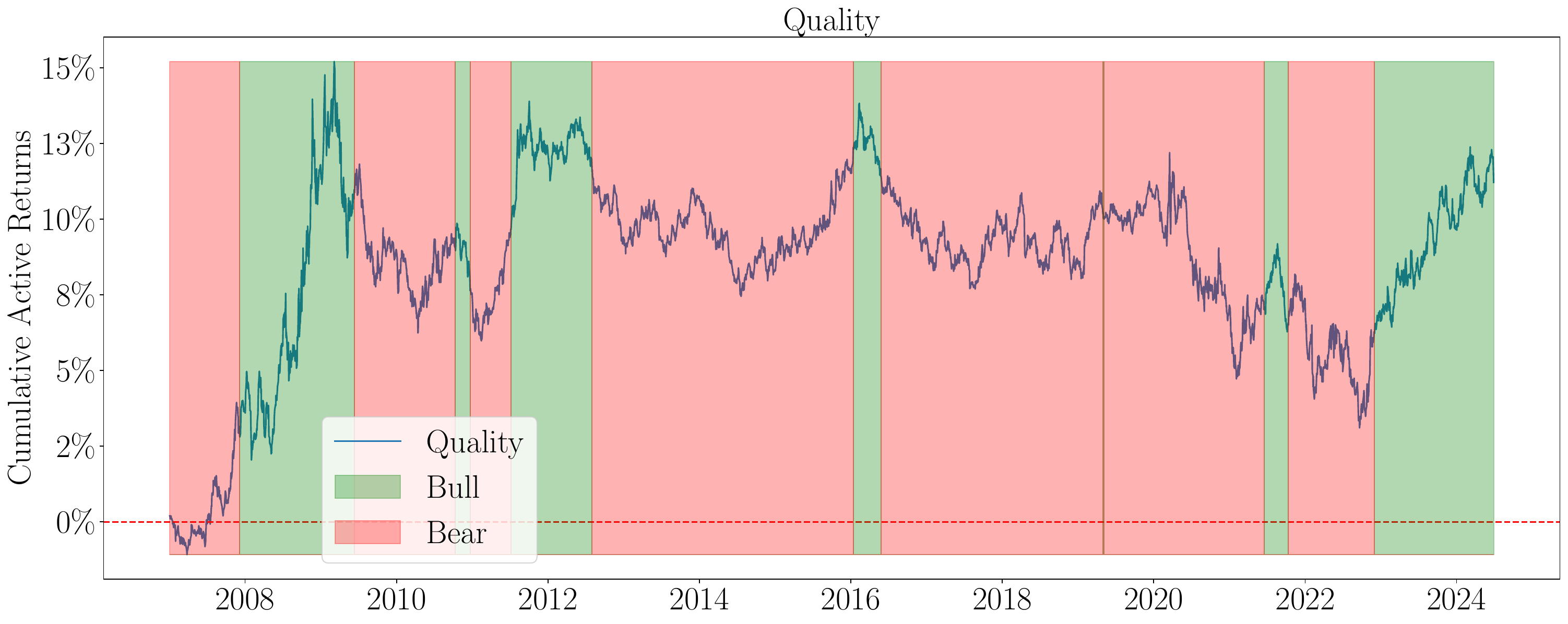}

    \includegraphics[width=\textwidth]{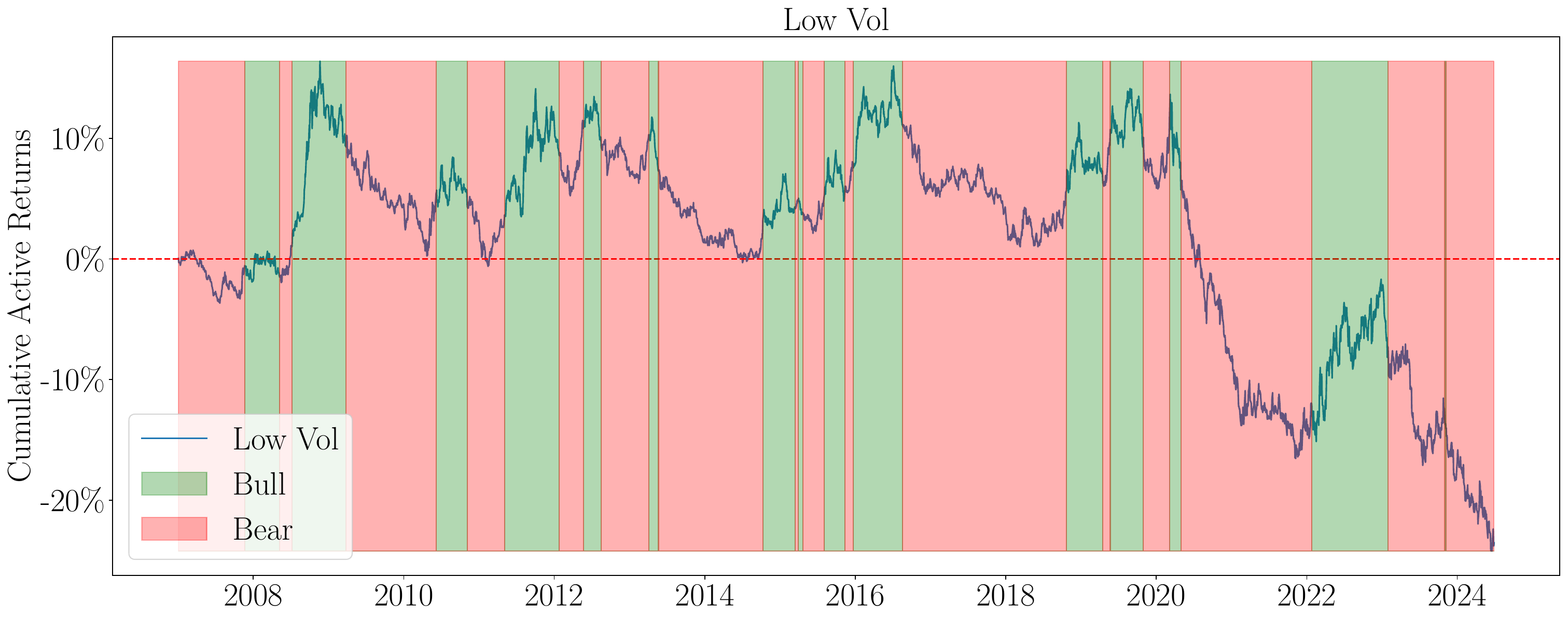}

    \includegraphics[width=\textwidth]{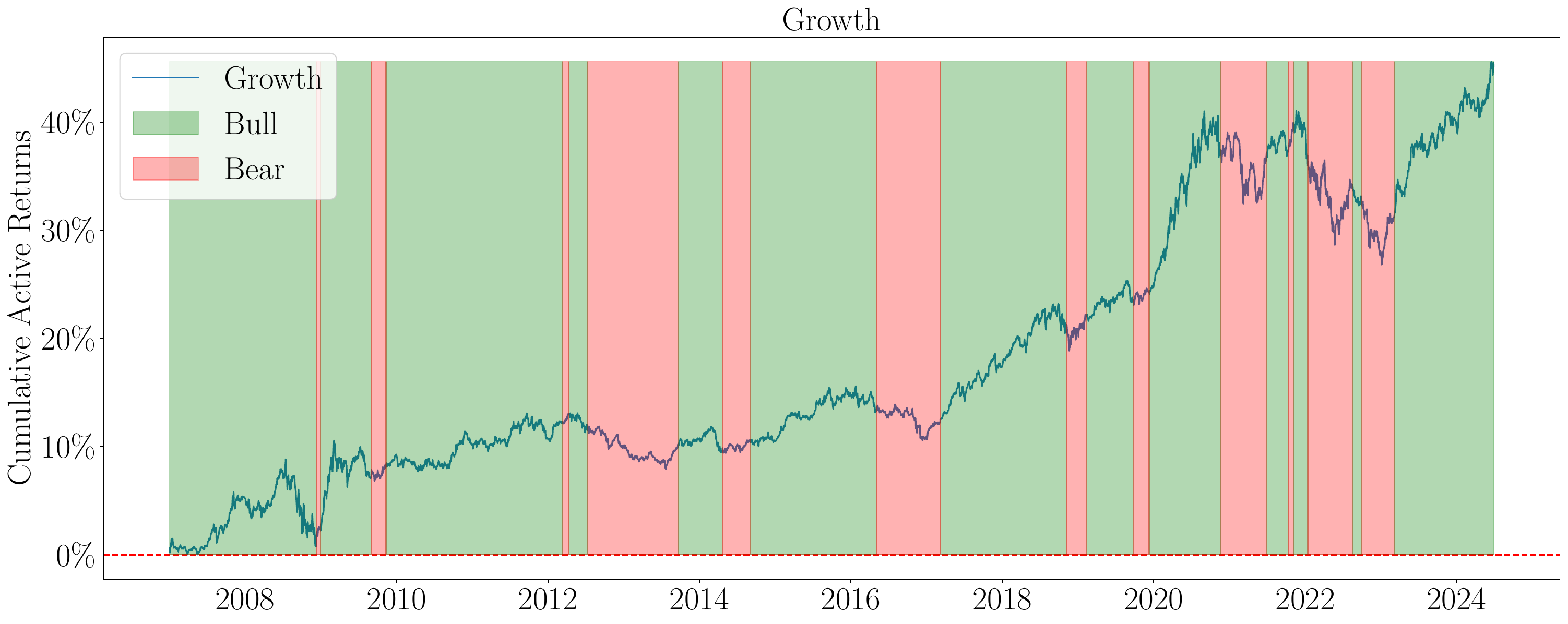}

    \vspace{2mm}

    \parbox{\textwidth}{\footnotesize
    Notes: The blue curve represents the cumulative factor active returns; the green and red shaded areas represent the online inferred bull and bear regimes after hyperparameter tuning, respectively, defined by whether the expected factor active return is positive or negative.
     }

    \caption{Online Inferred Regimes by the Sparse Jump Model (After Hyperparameter Tuning) Over the Test Period of 2007--2024 for Three Factors:  Quality, Low Volatility, and Growth}
\end{figure}

Exhibit~\ref{tab:single factor perf} presents the performance of the factor-specific long-short strategies for all factors, along with the correlations among these strategy returns over the testing period from 2007 to 2024.     %
The Sharpe ratios of these strategies range from a low of 0.16 for momentum to a high of nearly 0.4 for value and growth.
The varied Sharpe ratios reflect the differing levels of difficulty in accurately discerning bull and bear markets across factors. 
The frequency of regime shifts per year also varies. 
All Sharpe ratios are positive, demonstrating the potential utility of our regime-switching signal. 
The correlations among the factor-specific long-short strategies are generally low, below 0.5, suggesting potential for synthesis using the Black-Litterman model, which can effectively combine these factor-specific regime inferences while accounting for risk and portfolio constraints.

Exhibit~\ref{fig:oos regimes 1} displays the online inferred regimes after hyperparameter tuning using SJMs for the six factors, alongside their cumulative factor active returns.
These results complement the in-sample fitted regimes for the value factor under a specific hyperparameter setting, as shown in Exhibit~\ref{fig:value example}.
Overall, for most factors, the identified bear markets align with significant price declines, during which reducing market exposure helps avoid
losses and mitigate risk.
For factors where the long-short strategy achieves a higher Sharpe ratio, the identified regimes appear to be more accurate.
During periods of market oscillation, the SJM  produces rapidly flipping regimes, indicating its sensitivity to price reversals.    %
For the quality factor, which has particularly low volatility and max drawdown due to its defensive nature, we found the regime identification to be the least effective.

\section{Multi-Factor Portfolio Construction}      \label{sec:portfolio}

The final stage integrates all factor-specific regime inferences into a multi-factor portfolio. 
Here, the Black-Litterman model plays a central role in expressing our relative views on factor active performance into a \emph{long-only} and 100\% \emph{fully invested} allocation among the seven indices representing the market and six style factors. 
We emphasize our ``combination'' approach, which invests directly in pre-constructed factor indices.
Alternatively, \citet{Kolm2021factor} offer a framework for incorporating factor views into the Black-Litterman model to construct a ``bottom-up'' multi-factor portfolio from the security level.

Several alternative methods exist for synthesizing asset-specific regime inferences/forecasts. 
Though not explored here, a straightforward approach would be to employ any of the six constructed hypothetical long-short factor strategies individually or in combination (\mbox{e.g.,} through equal weighting, as in \citet{Bosancic2024}) as an overlay to an existing allocation to enhance returns. 
Alternatively, \citet{shu2024AA} investigate the use of traditional portfolio models, such as mean-variance and minimum-variance optimization, to synthesize asset-specific regime inferences into a dynamic allocation among asset classes.

The Black-Litterman model (see \citet{Kolm2021BL} for an overview) is highly relevant in the active management industry due to its emphasis on a benchmark portfolio. 
In our case, we use an equally weighted (EW) allocation among the seven indices, rebalanced quarterly,\footnote{
This rebalancing frequency aligns with that typically used by ETFs and mutual funds adopting EW strategies.
} as the benchmark. 
The robustness of this EW allocation as a benchmark in factor allocation contexts has been validated by \citet{Khang2023}. 
For prior expected returns, we use a risk aversion parameter of $\delta=2.5$, representing the ``world average risk tolerance'' \citep{He1999}, and estimate the covariance matrix using an exponentially weighted moving approach with a halflife of 126 days.

In constructing our view portfolios, we naturally express views on factor active returns through six relative portfolios, each comprising a 100\% long position on a factor and a 100\% short position on the market. 
The expected factor active return discussed in the previous section is used as the expected return for these view portfolios.
Under a specific confidence level,\footnote{
Using the same notation as in Appendix \hyperref[append:BL]{B}, the diagonal elements of the matrix $\bmOmega/\tau$ are chosen as a constant multiple of the diagonal elements of $\bmP\bmSigma\bmP\tran$. The inverse of this constant reflects the confidence level.
} we compute the posterior expected returns, which are then integrated into a standard mean-variance quadratic optimization with long-only and fully invested weight constraints to determine the optimal allocation.
Confidence levels are chosen to achieve a targeted tracking error relative to the benchmark, ranging from 1--4\%, allowing us to evaluate the performance of our dynamic allocation strategy.

\begin{table}[tb]
    \centering
\begin{tabular}{lrrrrrr}
\toprule
\multicolumn{7}{l}{\textbf{Panel A: Active Performance Relative to Market}}\\
\midrule
    & & Benchmark & \multicolumn{4}{c}{Dynamic Allocation}  \\
    \cmidrule(lr){3-3} \cmidrule(lr){4-7}   
 & Market & EW & $\text{TE}=1\%$ & $\text{TE}=2\%$ & $\text{TE}=3\%$ & $\text{TE}=4\%$ \\
\midrule
Active Return (\%) & 0.00 & 0.09 & 0.49 & 1.07 & 1.55 & 1.97 \\
Information Ratio & -- & 0.05 & 0.24 & 0.39 & 0.43 & 0.44 \\
Max Drawdown (\%)  & -- & $-$10.29 & $-$7.92 & $-$5.89 & $-$5.91 & $-$7.43 \\
\bottomrule
\\
\toprule
\multicolumn{7}{l}{\textbf{Panel B: Active Performance Relative to EW Benchmark}}\\
\midrule
    &  & Benchmark & \multicolumn{4}{c}{Dynamic Allocation}  \\
    \cmidrule(lr){3-3} \cmidrule(lr){4-7}   
 & Market & EW & $\text{TE}=1\%$ & $\text{TE}=2\%$ & $\text{TE}=3\%$ & $\text{TE}=4\%$ \\
\midrule
Active Return (\%) & $-$0.09 & 0.00 & 0.41 & 0.98 & 1.46 & 1.88 \\
Information Ratio & -- & -- & 0.40 & 0.49 & 0.49 & 0.47 \\
\bottomrule
\end{tabular}

    \vspace{2mm}

    \parbox{\textwidth}{\footnotesize Notes: 
    Our dynamic strategy maintains a long-only, fully-invested allocation among the market and six factor indices. 
``EW'' stands for the equally weighted portfolio of the seven indices, rebalanced quarterly.
``TE'' stands for the tracking error relative to the EW benchmark.
 A transaction cost of 5 basis points is applied for both buying and selling.  
 All values are annualized where applicable. 
 }

    \caption{Active Performance Analysis Relative to the Market and the EW Benchmark for Strategies Over the Test Period of 2007--2024}

    \label{tab:act perf}
\end{table}

\begin{figure}[tb]

    \begin{minipage}{\textwidth}
        \centering
    \begin{tabular}{l*{6}{r}}
\toprule
\multicolumn{7}{l}{\textbf{Panel A: Absolute Performance}}\\
\midrule
    &  & Benchmark & \multicolumn{4}{c}{Dynamic Allocation}  \\
    \cmidrule(lr){3-3} \cmidrule(lr){4-7}   
 & Market & EW & $\text{TE}=1\%$ & $\text{TE}=2\%$ & $\text{TE}=3\%$ & $\text{TE}=4\%$ \\
\midrule
Excess Return (\%) & 10.5 & 10.6 & 11.0 & 11.6 & 12.1 & 12.5 \\
Excess Risk (\%) & 20.1 & 19.6 & 19.4 & 19.2 & 19.1 & 19.1 \\
Sharpe Ratio & 0.52 & 0.54 & 0.57 & 0.60 & 0.63 & 0.65 \\
Max Drawdown (\%) & $-$54.9 & $-$52.9 & $-$52.5 & $-$52.2 & $-$51.4 & $-$50.5 \\
Turnover (\%) & -- & 3.8 & 237.2 & 395.5 & 522.5 & 627.5 \\
\bottomrule
\\
\toprule
\multicolumn{7}{l}{\textbf{Panel B: Cumulative Excess Returns of Strategies}}\\
\midrule
\end{tabular}
  \end{minipage}

    \begin{minipage}{\textwidth}
        \centering
        \includegraphics[width=\textwidth]{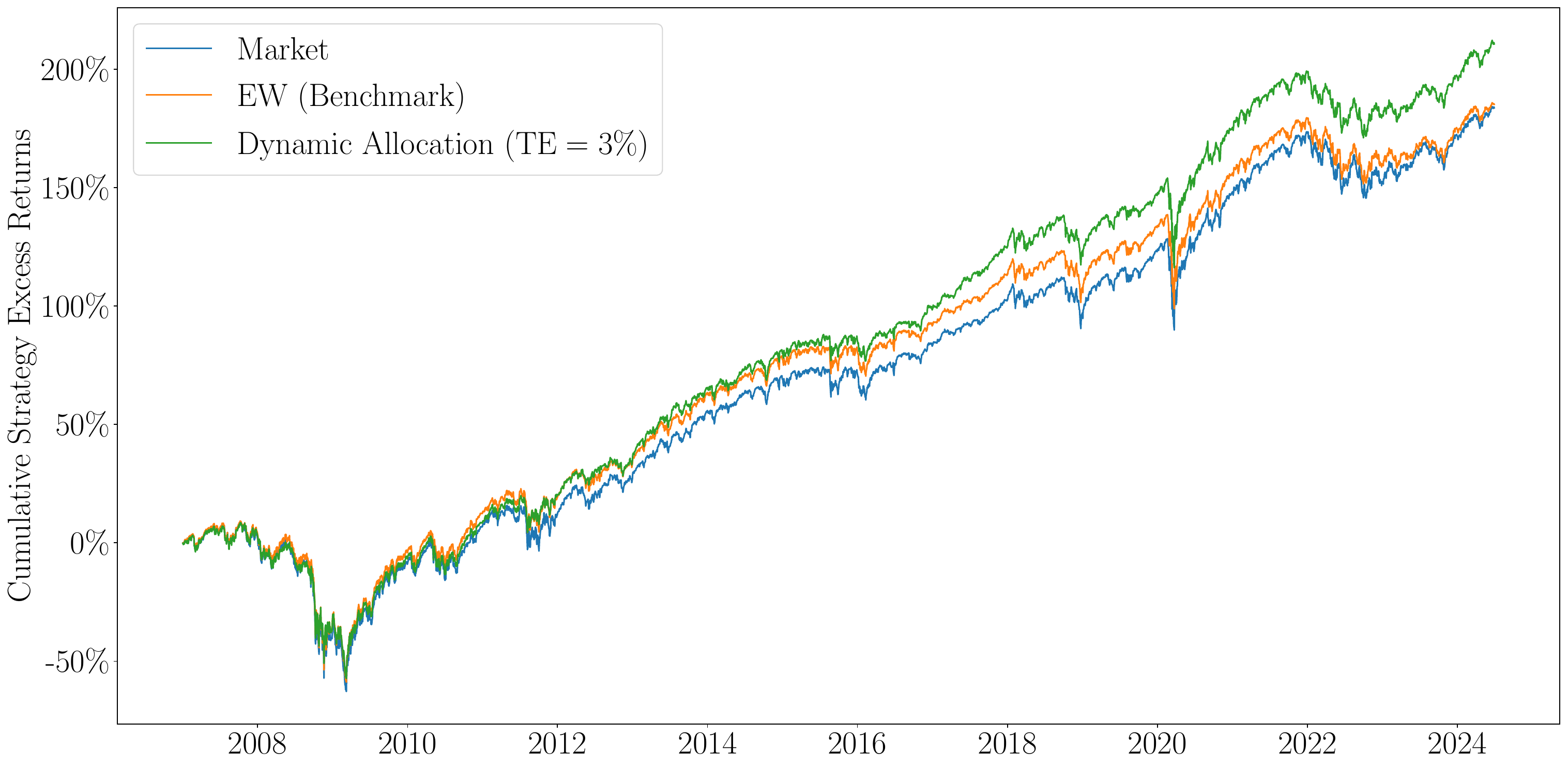}
    \end{minipage}

      \vspace{2mm}

    \parbox{\textwidth}{\footnotesize Notes: See the notes in Exhibit \ref{tab:act perf}.}

     \caption{Absolute Performance Analysis for Strategies Over the Test Period of 2007--2024}

   \label{tab:abs perf}
  
\end{figure}

Exhibit \ref{tab:act perf} presents the active performance of our dynamic multi-factor portfolio, measured relative to both the market and the benchmark EW portfolio, over the test period from 2007 to 2024. 
A transaction cost of 5 basis points is applied for both buying and selling throughout the performance evaluation.
In Panel A, all performance measures are shown relative to the market, with the EW benchmark exhibiting a negligible positive active return and information ratio (IR).
Our dynamic allocation significantly improves both active return and IR, with the IR increasing from 0.05 to a maximum of 0.44.
Additionally, the maximum drawdown relative to the market is reduced from $-$10\% to a low of $-$6\%.
In Panel B, when evaluating performance relative to the EW benchmark, our dynamic allocation achieves an IR of approximately 0.4 to 0.5, indicating robust outperformance over the EW benchmark.

Exhibit~\ref{tab:abs perf} presents our strategy's performance from an absolute perspective (\text{i.e.}, in excess of the risk-free rate).
The table shows that as the dynamic factor allocation allows for higher tracking error, both excess returns and the Sharpe ratio improve almost monotonically, while risk measures such as maximum drawdown decrease, albeit at the cost of higher turnover. 
An annualized one-way turnover of around 500\% remains within an acceptable range for active management and is reasonably implementable. 
One reason the max drawdown is not extensively mitigated, as seen in \citet{shu2024regime, shu2024AA, Bosancic2024}, is that we maintain a 100\% leverage, whereas those studies reduce leverage during identified bear markets to avoid significant wealth losses.
Incorporating an additional absolute view on the overall market's bull/bear regime, as developed in those studies, into our Black-Litterman model here could potentially achieve further risk mitigation.

In the accompanying figure, we plot the cumulative excess return curves for the market, the EW benchmark portfolio, and our dynamic allocation with a moderate tracking error of 3\%. 
Our strategy remains on par with the two baseline portfolios until around 2012, after which it steadily outperforms them through to the end of the period.
The empirical findings from both the factor-specific long-short strategy and the portfolio construction via the Black-Litterman model provide robust evidence of the potential to exploit factor cyclicality through regime analysis using the sparse jump model with a readily accessible feature set.

From the figure, it is notable that the outperformance of an equally weighted factor portfolio (benchmark) over the market steadily declined after about 2015, diminishing further following the COVID-19 market shock, and nearly disappeared post-2022 as the Federal Reserve implemented aggressive interest rate hikes.
Many well-known style factors have struggled recently\footnote{
As of August 2024, when the first draft of this article was written.
} due to a concentration of returns in a few of the largest companies in the U.S. equity market, amid a high-interest-rate environment not seen in nearly two decades. 
These factors have only begun to recover very recently, driven by expectations of future interest rate cuts and a rotation toward riskier assets.
It is beyond the scope and intention of our study and model to predict whether a comeback in factor investing is imminent.

\section{Conclusion} \label{sec:conclusion}

Motivated by the cyclical behavior of factor returns, this article proposes a dynamic factor allocation strategy through regime analysis based on each factor's active performance, synthesizing these factor-specific regime inferences using the Black-Litterman model.
A key distinction between factor regime analysis and factor timing strategies is that regime analysis does not aim to predict when a regime shift will occur but rather to identify when it has occurred, allowing investors to capitalize on changing risk-return dynamics, compared to a static benchmark allocation.

We employ the sparse jump model (SJM) as the regime identification model, which clusters temporal features while imposing a penalty for each transition in the hidden state sequence and assigns different weights to features based on their in-sample clustering effect.
The input features include return and risk measures calculated from historical factor active returns, as well as features representing the general market environment. 
The regimes identified by the SJM demonstrate improved robustness and accuracy compared to traditional regime-switching models.

Our empirical study focuses on a U.S. equity investment universe comprising seven long-only indices representing the market and six style factors -- value, size, momentum, quality, low volatility, and growth.
We construct a hypothetical single-factor long-short strategy to evaluate the quality of the factor-specific regime inferences and tune hyperparameters based on its performance.
This strategy also closely mirrors the active weights in the Black-Litterman model when an active view of a factor is applied.
We achieve positive Sharpe ratios across all factors, with low correlations among them, demonstrating the high accuracy of our regime inferences.

Finally, we apply the Black-Litterman model, using the equally weighted allocation as the benchmark, to determine the optimal dynamic allocation derived from the factor-specific regime inferences. 
Empirical results show that our dynamic allocation strategy outperforms a static approach, enhancing key metrics such as the information ratio and Sharpe ratio while reducing maximum drawdown. These findings underscore the effectiveness of leveraging regime-switching signals to exploit factor cyclicality and improve factor allocation performance.

Going forward, the statistical jump model framework can be enhanced through alternative approaches for measuring the (dis-)similarity between a given time period and the cluster centroid. 
For example, advanced neural network structures such as variational auto-encoders (VAEs) have proven beneficial in intraday trading by developing regime-aware execution systems \citep{Sawhney2021}. 
Moreover, \citet{fantulin2024} have proposed a JM-inspired recurrent neural network (RNN)-based model for regime identification. 
Another promising extension is the use of generative models to expand the training and validation datasets, which could help mitigate the challenge of overfitting -- a constant issue in the field of investment.

\clearpage

\appendix

\section*{Appendix A}   \label{append:data}

\subsection*{Factor Indices}

Exhibit \ref{tab:data details} provides a detailed overview of the factor indices, including the number of stock holdings (as of mid-2024) and the tickers of the actively traded index-tracking ETFs used in our study.

\begin{table}[htb]
\centering
\begin{tabular}{cccc}
\toprule
\textbf{Factor} & \textbf{Index Name} & \makecell[c]{\textbf{Number} \\ \textbf{of Stocks}} & \makecell[c]{\textbf{ETF} \\ \textbf{Ticker}} \\ 
\midrule
Market & \makecell[c]{MSCI USA Index} & 601   &   PBUS \\  
\midrule
Value & \makecell[c]{MSCI USA Enhanced Value   Index\tablefootnote{
Replaced by the MSCI USA Value Weighted   Index prior to 1997/12.
}
} & 150 & VLUE \\  
\midrule
Size & \makecell[c]{MSCI USA Low Size   Index\tablefootnote{
Replaced by the MSCI USA Equal Weighted   Index prior to 1994/6.
}
} & 601 & SIZE  \\
\midrule
Momentum & \makecell[c]{MSCI USA Momentum  SR Variant Index\tablefootnote{
Replaced by the MSCI USA Momentum  Index prior to 2002/6.
}
} & 125 & MTUM \\
\midrule
Quality & \makecell[c]{MSCI USA Sector Neutral Quality  Index\tablefootnote{
Replaced by the MSCI USA Quality   Index prior to 1998/12.
}
} & 125 & QUAL \\
\midrule
Low Volatility & \makecell[c]{MSCI USA Minimum Volatility  Index\tablefootnote{
Replaced by the MSCI USA Risk Weighted   Index prior to 1994/6.
}
} & 170 & USMV  \\
\midrule
Growth & \makecell[c]{Russell 1000 Growth  Index} & 440 & IWF \\
\bottomrule
\end{tabular}
\caption{Detailed Overview of Factor Indices, Number of Stock Holdings, and Index-Tracking ETF Tickers}
\label{tab:data details}
\end{table}

\section*{Appendix B}   \label{append:BL}

\subsection*{Details of the Black-Litterman Model}

Most of the results can be found in \citet{He1999}. 
We present them here as they provide significant motivation for our method of constructing the factor-specific long-short strategy.

Consider $N$ securities and $K$ view portfolios, represented by $\bmp_1,\ldots,\bmp_K \in \R^N$, assembled into the matrix $\bmP := \begin{bmatrix}\bmp_1\tran\\\cdots\\\bmp_{K}\tran\end{bmatrix} \in \R^{K\times N}$.
Let $\bmv \in \R^K$ denote the expected returns of the view portfolios, and $\bmw^{\text{bmk}} \in \R^N$ the benchmark portfolio weights.
If the investor optimizes a quadratic mean-variance utility function with no constraints, the optimal portfolio weights in the Black-Litterman model are given by:
\begin{equation}
    \bmw^{\text{BL}} = \bmw^{\text{bmk}} + \bmP\tran\bmlambda,
\end{equation}
where $\bmlambda = \delta^{-1}\lp{\bmP\bmSigma\bmP\tran + \bmOmega/\tau}^{-1}\lp{\bmv - \bmP\bmpi}$. Here, $\delta$ is the risk aversion parameter, $\bmSigma \in \R^{N\times N}$ is the covariance matrix estimate, and $\bmOmega \in \R^{K\times K}$ and $\tau$ denote the uncertainty matrix and parameter in the views and the prior, respectively.
$\bmOmega$ is a diagonal matrix with diagonal elements $\omega_1,\ldots,\omega_K$.

Thus, the active weights $\bmP\tran\bmlambda = \sum_j \lambda_j \bmp_j$ represent a linear combination of all the view portfolios, weighted by the vector $\bmlambda$. 
Using the Schur complement, the explicit formula for each element $\lambda_j$ is given by:
\begin{equation}
\lambda_j = \frac{1}{\delta} \frac{v_j - \bmp_j\tran\bmmu^{\text{BL}}_{-j}}{\eta_j},
\end{equation}
where $\bmmu^{\text{BL}}_{-j}$ denotes the posterior expected return estimation when all but the $j$-th view are considered, and $\eta_j$ adjusts for risk:
\begin{equation*}
    \eta_j = \bmp_j\tran\bmSigma\bmp_j + \omega_j/\tau - \bmp_j\tran\bmSigma\bmP\tran\lp{\bmP\bmSigma\bmP\tran + \bmOmega/\tau}^{-1}\bmP\bmSigma\bmp_j.
\end{equation*}

Since we aim to tune the hyperparameters for each factor individually, $\lambda_j\bmp_j$ is a natural choice for constructing the hypothetical long-short strategy. 
Furthermore, the weight $\lambda_j$ is approximately proportional to the expected return $v_j$ for the active return, assuming the risk $\eta_j$ remains stable over time and the impact of the other views, expressed in $\bmp_j\tran\bmmu^{\text{BL}}_{-j}$, is minimal.
We cap the estimated expected active return at $\pm 5\%$ per annum to mitigate the impact of extreme values.

{
\small

\bibliographystyle{apalike}
\bibliography{lit_factor}

}

\end{document}